\documentclass{PoS}

\usepackage{amsmath}

\title{Challenges Implementing non-Abelian $SU(2)$ Quantum Chromodynamics Gauge Links On a Universal Quantum Computer}

\ShortTitle{Implementing $SU(2)$ Gauge Links on a Quantum Computer}

\author{\speaker{Patrick Dreher}
        North Carolina State University\\
        E-mail: \email{padreher@ncsu.edu}}

\abstract{The traditional approach for studying the physics of the strong interactions employs a basic computational construct originally proposed by Wilson in the 1970s.  Over the years additional enhancements have been added to this formulation to improve computational performance and accuracy.  This formulation has been successfully implemented on high performance computing systems and has yielded accurate calculations for many static properties of the strong interactions (such as the hadron mass spectrum). With the recent advances in quantum computing, the question that is now being asked is whether an equivalent type of gauge invariant formulation of a field theory can be constructed on a quantum computer to calculate dynamical processes that cannot be simulated on a traditional supercomputer. Using the Quantum Link Model (QLM) plus the concept of rishons, this paper will specifically focus on the challenges implementing a basic gauge link lattice construct using $SU(2)$ non-Abelian links for illustration.  The paper will also discuss the physics that may potentially be simulated on a quantum computer with this construct and speculate on the prospects for having quantum computers become a part of the set of hardware platforms for lattice gauge theory simulations in the future.}

\FullConference{The 36th Annual International Symposium on Lattice Field Theory - LATTICE2018\\
		22-28 July, 2018\\
		Michigan State University, East Lansing, Michigan, USA.}

\begin{document}

\section{Introduction}
The development of supercomputers has allowed physicists to numerically study several characteristics of the strong interactions (such as the hadron mass spectrum). However, computational studies of many strong interaction dynamical processes such as the real-time evolution of heavy-ion collisions or the phase structure of dense QCD matter cannot be simulated with digital supercomputers because of the inherent sign problem in the equations that manifests itself in simulations.  Similar types of sign problem challenges are also present in many problems in condensed matter physics.  Recent advances in the development of new computational architectures based on the principles of quantum mechanics rather the present base-2 digital computer architectures have opened the possibility that these new systems may be able to make progress in areas that cannot be handled with conventional computers.

Within the fields of nuclear and high energy physics, no one has been able to provide analytical solutions to the physics equations of the strong interactions  because of their complexity.  Consequently, physicists must rely on lattice Quantum Chromodynamics (LQCD) computational simulations to make progress in these areas.  A viable quantum computer would provide a welcome additional computational hardware tool.

At the present time the Standard Model is the best theory available that describes the physics of the strong interactions.  To simulate these QCD equations Ken Wilson proposed a computational procedure \cite{Wilson1974} that placed the strongly interacting particles (quarks) on a lattice of points representing a four-dimensional space-time.  These points are connected by link matrices $U_{xy}$ representing the gluon fields which are the carriers of the strong force. In this formulation the links act as gauge invariant parallel transporters connecting the strong force between the quarks on these lattice points with the $U_{xy}$ link variables represented by continuous gauge groups.   While this type of Quantum Chromodynamics formulation can be implemented on conventional computers, such mathematical constructs are not compatible with quantum computing architectures.

This paper will focus on the specific challenges of re-formulating the traditional digital computational version LQCD so that it can be implemented on a quantum computing system. Although there are several distinct quantum computing computational efforts underway (ex. optical lattices, etc. \cite{Cold-atom-2013, Cold-atom-2015}) this paper will focus on applying QCD constructs to publicly accessible superconducting transmon devices such as the IBM Q quantum computing system \cite{IBM}.

The paper is organized as follows.  Section \ref{SEC:QLM} summarizes the essential properties of the Quantum Link Model (QLM) and outlines a gauge invariant representation that factorizes the QLM operators into rishon components on the ends of the quantum links.  Section \ref{SEC:Challenges} discusses the challenges of applying $SU(2)$ gauge group matrices onto the gauge links between the lattice points in a way that they can be implemented on a universal quantum computer and finally Section \ref{SEC:Summary} summarizes next steps.

\section{The Quantum Link Model}
\label{SEC:QLM}
Beginning in the 1980s various researchers began experimenting with developing alternate Quantum Chromodynamics formulations from the traditional approach to try improving the performance and accuracy of numerical lattice based simulations.  These alternatives proposed modifying the real and imaginary parts of the traditional representation of the link variables ($U_{xy}$ matrices) connecting the lattice sites $x$ and $y$.  The proposed change modified the link variables from the traditional complex valued matrices to ones where the link variables are now intrinsically quantum mechanical objects.  This idea is similar to the generalized quantum spins that are represented by non-commuting Hermitian operators.  This overall approach was named the Quantum Link Model (QLM).

\subsection{Early Development of Alternatives to Traditional QCD Implementations}
In 1981  Horn \cite{Horn1981} published such a model for the gauge groups $U(1)$ and $SU(2)$.  In the early 1990s Orland and Rohrlich \cite{OandR-1990, O-1992} used these basic ideas to describe gauge magnets.  In the late 1990s Chandrasekharan and Wiese, \cite{CandW-1997} along with Schlittgen and Wiese \cite{SandW2000} drew on the analogy that ordinary spin systems can be related to quantum spin models and connected the ordinary lattice gauge theories to these non-commuting operators to develop an alternative regularization of non-Abelian gauge theories.

For problems in particle physics, these gauge fields on the links have dynamical quantum degrees of freedom and are not just classical background fields.  By representing the link matrices as non-commuting operators acting in a Hilbert space applied in a QLM formulation,  Brower, et.\ al.\ \cite{brower1999} used these QLM ideas to develop an $SU(3)$ lattice formulation of QCD. This formulation had continuous gluon fields emerging via dimensional reduction from the collective dynamics of the $(4+1)$-dimension discrete quantum link variables with the quarks identified as domain wall fermions at the edge of the extra dimension.  This demonstrated that when these models are extrapolated to the continuum limit for $SU(3)$ using dimensional reduction techniques, the standard strong interaction theory of QCD is recovered with chiral quarks as domain wall fermions.  In 2004 the work was extended to include other non-Abelian gauge groups such as $SU(N)$, $U(N)$, $SO(N)$, and $Sp(N)$, and the exceptional group $G(2)$ \cite{brower2004}.

\subsection{The Quantum Link Model and Quantum Computing}
The starting point for formulating QCD on a quantum computer requires that the physics should be expressed in a Hamiltonian formulation.  Kogut and Susskind \cite{K-S-1975} constructed a generalized lattice QCD Hamiltonian that can be written as
\begin{eqnarray}
\label{K-S-Hamiltonian}
H&=&- t \sum_{\langle x y \rangle}
\left(s_{xy} \psi_x U_{xy} \psi_y + \mathrm{h.c.}\right) +
m \sum_x s_x \psi_x^{i \dagger} \psi_x^i \nonumber \\
&+&\frac{g^2}{2} \sum_{\langle x y \rangle}
\left(L_{xy}^a L_{xy}^a + R_{xy}^a R_{xy}^a\right) +
\frac{{g'}^2}{2} \sum_{\langle x y \rangle} E_{xy}^2 \nonumber \\
&-&\frac{1}{4 g^2}
\sum_{\langle w x y z \rangle} \left(\mathrm{Tr} U_{wx} U_{xy} U_{yz} U_{zw} +
\mathrm{h.c.}\right) \nonumber \\
&-& \gamma \sum_{\langle x y \rangle}
\left(\mathrm{det} U_{xy} + \mathrm{h.c.}\right).
\end{eqnarray}
In this equation the $U_{xy}$ terms are $N \times N$ matrices (with matrix elements $U_{xy}^{ij}$) associated with the oriented link connecting the neighboring points $x$ and $y$.  $E_{xy}$, $L_{xy}^a$, and $R_{xy}^a$ are Abelian and non-Abelian electric field operators associated with the link $\langle x y \rangle$, and $g'$ and $g$ are the corresponding Abelian and non-Abelian gauge couplings. The magnetic field energy is connected with the elementary plaquette $\langle w x y z \rangle$.  The term proportional to $\gamma$ explicitly breaks a $U(N)$ gauge symmetry down to $SU(N)$.  (For completeness, the other terms such as $\psi^{i \dagger}_x$ and $\psi^i_x$ where $i \in \{1,2,\dots,N\}$ are the fermion creation and annihilation operators that obey standard anti-commutation relations.  The $"s"$ terms are sign factors associated with the detailed construction of staggered fermions on this $d$-dimensional spatial lattice coupled to a $U(N)$ gauge fields represented by the $U_{xy}^{ij}$ matrix elements.)

It can be shown that the $E_{xy}$, $L_{xy}^a$, and $R_{xy}^a$ operators in the Hamiltonian obey the commutation relations
\begin{eqnarray}
\label{linkalgebra-1}
&&[L^a,L^b] = 2 i f_{abc} L^c, \ [R^a,R^b] = 2 i f_{abc} R^c, \nonumber \\
&&[L^a,R^b] = [E,L^a] = [E,R^a] = 0, \nonumber \\
&&[L^a,U] = - \lambda^a U, \ [R^a,U] = U \lambda^a, \ [E,U] = U.
\end{eqnarray}
From these commutation relations (Eq.\ \ref{linkalgebra-1}) it follows that operators associated with different links commute with each other. The $\lambda$ terms in Eq.\ \ref{linkalgebra-1} are connected to the Hermitian generators of $SU(N)$
\begin{equation}
\label{group-theory}
[\lambda^a,\lambda^b] = 2 i f_{abc} \lambda^c, \
\mbox{Tr} \lambda^a \lambda^b = 2 \delta^{ab},
\end{equation}
where $f_{abc}$ are the structure constants of the $SU(N)$ algebra (Eq.\ \ref{group-theory}).

A key observation to note is that in the traditional Kogut-Susskind lattice QCD Hamiltonian the link matrices $U_{xy}$ can assume continuous values in the gauge group $U(N)$ or $SU(N)$ and the $E_{xy}$, $L_{xy}^a$, and $R_{xy}^a$ produce the corresponding canonically conjugate momentum operators that are the derivatives with respect to the matrix elements of $U_{xy}$.  The commutation relations that result from this type of construct yield an infinite-dimensional Hilbert space per link.  Quantum computers need a finite-dimensional Hilbert space representation to model strong interaction physics problems.  As a result, the traditional QCD formulation is inadequate for implementation on quantum computers. Progress with QLM research \cite{Wiese-2014a, Wiese-1211, Wiese-1712} shows promise as an alternative QCD formulation for implementing strongly interacting physics problems on a quantum computer.

\subsection{Rishons}
It has been recognized from the outset that the properties of the link matrices are a critical factor in determining the types of modifications that will be needed to transform the gauge links from an infinite to finite dimensional Hilbert space representation.  Calculating and analyzing the generators of the algebra provides a critical piece of information toward that goal.

For a link variable represented by a matrix reflecting the group properties of the strong interactions the real and imaginary parts of the $N^{2}$ matrix elements $U_{xy}^{ij}$ yields $2N^{2}$ Hermitian generators.  The operators $L^{a}$ and $R^{a}$ represent the $SU(N)_{L}$ x $SU(N)_{R}$ gauge transformations on the left and right ends of the gauge link and contribute another $2N^{2}-1$ generators.  The operator $E$ provides one more generator. The aggregated total of $4N^{2}-1$ generators forms an $SU(2N)$ algebra. This is the smallest algebra under which the commutation relations in Eq.\ \ref{linkalgebra-1} close.

The Quantum Link Model allows these quantum operators to be expressed as bilinears of fermionic creation and annihilation operators and have been given the name rishons. These rishon creation and annihilation operators can be defined as $c^{i\dagger}_{\pm}$ and $c^{i}_{\pm}$,.  The $\pm$ indicate the left and right ends of the link and the $i \in \{1,2,...,N\}$ is the color index.  These creation annihilation operators obey standard anti-commutation relations
\begin{eqnarray}
&& \{c^i_{x,\pm k},c^{j \dagger}_{y,\pm l}\} = \delta_{xy} \delta_{\pm k,\pm l} \delta_{ij},  \nonumber \\
&& \{c^i_{x,\pm k},c^j_{y,\pm l}\} = \{c^{i \dagger}_{x,\pm k},c^{j \dagger}_{y,\pm l}\} = 0.
\end{eqnarray}

The $SU(2N)$ generators obey the commutation constraint of Eq.\ \ref{linkalgebra-1} and produce a finite number of independent gauge transformations on the left and right ends of the link. The $L^a_{xy}$, $R^a_{xy}$ and $E_{xy}$ terms can be expressed in terms of rishon creation and annihilation operators in the form
\begin{eqnarray}
\label{LRE}
&& L^a_{xy} = c^{i \dagger}_{x,+} \lambda^a_{ij} c^j_{x,+}, \
 R^a_{xy} = c^{i \dagger}_{y,-} \lambda^a_{ij} c^j_{y,-}, \nonumber \\
&& E_{xy} = \frac{1}{2}(c^{i \dagger}_{y,-} c^i_{y,-} - c^{i \dagger}_{x,+} c^i_{x,+})
\end{eqnarray}
An additional important result from this representation is that the link matrices $U^{ij}_{xy}$ are now directly connected to the rishon operators at the ends of each link (Eq.\ \ref{U-rishon}).
\begin{equation}
\label{U-rishon}
U^{ij}_{x,y} = c^i_{x,+} c^{j \dagger}_{y,-}.
\end{equation}
All operators that have been introduced as part of these modifications (including the Hamiltonian) commute with the rishon number operator ${\cal N}_{xy} = c^{i \dagger}_{y,-} c^i_{y,-} + c^{i \dagger}_{x,+} c^i_{x,+}$ on each individual link.  This only then requires one to consider a fixed rishon number for each link.  Essentially this is equivalent to working in an given irreducible representation of $SU(2N)$.

\section{$SU(2)$ non-Abelian Gauge Links on a Quantum Computer}
\label{SEC:Challenges}
The analysis from Section \ref{SEC:QLM} can now be applied to the specific illustration for $SU(2)$ non-Abelian gauge links.  Following the analysis of \cite{SandW2000} and the discussion regarding the rishon representation of the $SU(2)$ Quantum Link Model in \cite{Kagome}, it can be shown that the electric flux operators on the xy link at the left and right ends of a gauge link preserve the commutation relations of Eq.(\ref{linkalgebra-1}) and can be written as
\begin{align}
\label{lambda}
&L^a_{xy} = \frac{1}{2} {c^i}_{xy,-}^\dagger \sigma^a_{ij} c^j_{xy,-}, \nonumber\\
&R^a_{xy} = \frac{1}{2} {c^i}_{xy,+}^\dagger \sigma^a_{ij} c^j_{xy,+},
\end{align}
In this case the $\lambda$ terms in Eq.\ \ref{LRE} are replaced by the $SU(2)$ Pauli matrices $\sigma$.  The $SU(2)$ matrix elements $U_{xy}^{ij}$ in Eq.\ \ref{U-rishon} can now be expanded into components
\begin{align}
&U_{xy}^{11}={c^1}_{xy,+}^\dagger c^1_{xy,-} + {c^2}_{xy,-}^\dagger c^2_{xy,+}, \nonumber \\
&U_{xy}^{12}= {c^2}_{xy,+}^\dagger c^1_{xy,-} - {c^2}_{xy,-}^\dagger c^1_{xy,+}, \nonumber \\
&U_{xy}^{21}={c^1}_{xy,+}^\dagger c^2_{xy,-} - {c^1}_{xy,-}^\dagger c^2_{xy,+}, \nonumber \\
&U_{xy}^{22}={c^2}_{xy,+}^\dagger c^2_{xy,-} + {c^1}_{xy,-}^\dagger c^1_{xy,+}.
\end{align}
where $i$ and $j$ are $SU(2)$ color indices, and $-$ and $+$ refer to the $x$ and $y$ ends of the link $\langle xy \rangle$.  From the equations it is noted that the quantum link operator $U_{xy}$ shuffles a rishon from one end of the link to the other.  The number operator ${\cal N}_{xy} = {c^i}_{xy,+}^\dagger c^i_{xy,+} + {c^i}_{xy,-}^\dagger c^i_{xy,-}$ keeps the total number of rishons per link fixed.

In the QLM  vector representation for $SU(2)$ there are ${\cal N}_{xy} = 2$ rishons per link.  These are color-doublet fermions residing at the ends of a link and obey standard anti-commutation relations.  There are 4 possible combinations where one rishon can reside at each end of the link.  There are also 2 possible combinations where a pair of rishons can reside at the same end of the link and form a color-singlet.  Due to the rishon's fermionic nature, when two of them  sit on the same end of the link they necessarily form a color-singlet.  In this case, one possibility is a symmetric superposition of a two-rishon singlet sitting on the left and on the right end of the link.  The other possibility is an anti-symmetric superposition (which does not contribute to the QLM dynamics).

Each gauge link matrix $U_{xy}$ connects two nearest neighbor lattice sites $x$ and $y$.  Every $SU(2)$ link matrix on each gauge link has a 2 rishon configuration in one of these 6 states.  (Models with a larger number of dimensions and/or higher order gauge groups (such as $SU(3)$ for modelling standard QCD) require 20 qubits per gauge link $SU(3)$.)  The left and right rishon components of the gauge link $U_{xy}$ attach to the lattice site corresponding to the left ($x$) and right ($y$) sides of the gauge link.  There are additional adjacent gauge links that also attach to the two lattice sites $x$ and $y$.  Connecting these adjacent links at each lattice and properly accounting for their fermionic nature requires that the rishons sharing the common lattice site "$x$" (without external charges) must form an overall configuration that obeys Gauss's Law.

In this rishon formulation, the minimum number of qubits required to program an $SU(2)$ gauge link onto a quantum computer requires 3 qubits ($|\alpha\beta\gamma>$). Qubit $\alpha$ assigns a $"0"$ if it is a rishon creation operator or a $"1"$ for a rishon annihilation operator.  Qubit $\beta$ assigns states to the $x$ and $y$ ends of the link $\langle xy \rangle$.   A $"0"$ is assigned to qubit 2 if the operator is on the left side of the link ($x$ side represented by a $-$) or a $"1"$ if the operator is on the right side of the link ($y$ represented by a $+$) .  Qubit $\gamma$ is assigned a $"0"$ for the first color index $i$ and a $"1"$ for the second color index $j$ of the $SU(2)$ matrix. Using this configuration, a first calculation of $SU(2)$ using QLM and rishons on a quantum computer might consider a 12 qubit plaquette $P=U_{xy}U_{yz}U_{zw}U_{wx}$ at different couplings on a 2-d lattice with $SU(2)$ gauge links.

\section{Summary and Next Steps}
\label{SEC:Summary}
Quantum computers such as the superconducting transmon devices are at the beginning stages of development.  At the present time the largest IBM Q available has 20 qubits.  A 50 qubit IBM Q machine may become available sometime in 2019.  QLM calculations with $SU(2)$ are possible today but $SU(3)$ simulations are beyond the capacity of an IBM Q type quantum computer at this time.

Although the universal quantum computer approach for standard QCD may not be fully implementable on quantum computing hardware at the present time, there are discussions in the quantum computing community that the first practical quantum computers may be hybrids of both a conventional and quantum machine.  Certainly the lessons learned from these early investigations of re-formulating the physics of the strong interactions in a way that is compatible with a quantum computer are valuable and can contribute to the designs of hybrid architectures where a portion of the physics problem is simulated on a conventional machine and specific sections of the code that are amenable to quantum computing speedup are allocated to the quantum computer.

\section{Acknowledgments}
The author gratefully acknowledges partial support for this work through an IBM Faculty Award and also is appreciative of some QLM conversations with Uwe-Jens Wiese.

\end{document}